# Vision and Challenges for Knowledge Centric Networking (KCN)

Dapeng Wu, *Fellow, IEEE*, Zhenjiang Li, *Member, IEEE*, Jianping Wang, *Member, IEEE*, Yuanqing Zheng, *Member, IEEE*, Mo Li, *Member, IEEE*, Qiuyuan Huang, *Member, IEEE*

*Abstract*—In the creation of a smart future information society, Internet of Things (IoT) and Content Centric Networking (CCN) break two key barriers for both the front-end sensing and back-end networking. However, we still observe the missing piece of the research that dominates the current networking traffic control and system management, *e.g.,* lacking of the knowledge penetrated into both sensing and networking to glue them holistically. In this paper, we envision to leverage emerging machine learning or deep learning techniques to create aspects of knowledge for facilitating the designs. In particular, we can extract knowledge from collected data to facilitate reduced data volume, enhanced system intelligence and interactivity, improved service quality, communication with better controllability and lower cost. We name such a knowledge-oriented traffic control and networking management paradigm as the Knowledge Centric Networking (KCN). This paper presents KCN rationale, KCN benefits, related works and research opportunities.

## I. Introduction

We have recently witnessed the proliferation of two emerging technologies that are evolving our sensing, computing and networking capabilities, to enable the vision of a smart future information society: Internet of Things (IoT) [3] and Content Centric Networking (CCN) [7]. The IoT technology strives to fundamentally advance the periphery of sensing and enables the scene that "every" physical object can be sensed, so that intelligent interactions between users and the digital space could be viable to fulfill user's intents. Aligning with this trend, the CCN technology [3] is further proposed to augment the underlying networking services for the data exchange, tailored for the data-centric feature. CCN leverages in-network cache and allows end users to obtain data from anywhere in the network, instead of end sensing devices.

These two technologies together overcome two key barriers for both front-end *sensing* and back-end *networking* in the creation of a smart information society for various domains. However, we still observe a daunting challenge that remains unsolved to limit the further advance of the design. User's intents to interact with the digital space, *e.g.,* user's objectives and desired system feedback, can be highly diversified and even time-varying. Lacking the tailored solution to derive from the front-end sensing to the anticipated feedback from the system, existing network designs usually require to deliver all necessary sensing data and in-network traffic to the end users and further realize the missing intelligence on the end devices.

1) However, with the increasing number of users involved, big data, as a result, is a natural consequence, yet much of the traffic can be highly redundant or even unnecessary. Even if networking infrastructures have rich resources, the rapidly increased data volume could incur harmful burden and rapidly overwhelm system's computing and networking capabilities.

2) Due to the high traffic demand, networking systems tend to become more sophisticated in the operation, which may even need to be further adjusted according to the application's varying context. The traditional system control and networking management, mainly based on human's intelligence and experiences, will thus not be sufficient and flexible enough.

3) Implementing all the missing intelligence to end users will inevitably increase the bar on the requirements of the end device, *e.g.,* capable CPUs for intensive computing tasks.

This paper finds that the emerging machine or deep learning techniques have the great potential to tackle above challenges. The key insight is applying the learning techniques to derive various in-network knowledge tailored for aspects of desired networking traffic control and system management designs:

- *knowledge creation*: using learning techniques to extract descriptive knowledge from raw IoT or edge sensory data, which could distill the raw data to create valuable knowledge and minimize the data volume to be transmitted.
- *knowledge composition*: further deriving automatically-learned rules to produce useful information and feedback to fit user's demands, which provides an in-network digestion of user's intents to interact with the system.
- *knowledge distribution*: efficiently distributing the generated knowledge and information to both the end users and desired devices for operations, to avoid blindly delivering all sensing data and traffic to the end users.

Such a new paradigm advances the traditional traffic control and networking system design viewed from a communication-oriented perspective to the knowledge-oriented one, which is named as knowledge centric networking (KCN) in this paper. KCN leverages the in-network computing, in-network storage, and in-network communication, which are available in current network stacks, to create knowledge in need. Machine learning is capable of exploiting the hidden relationship from voluminous input data to complicated system outputs and further adapting the learning results in the new environments

Dapeng Wu is with Department of Electrical and Computer Engineering, University of Florida, FL 32611, USA. Correspondence author: Dapeng Wu, Email: dpwu@ieee.org. Zhenjiang Li and Jianping Wang are with Department of Computer Science, City University of Hong Kong, Hong Kong; Emails: zhenjiang.li@cityu.edu.hk, jianwang@cityu.edu.hk. Yuanqing Zheng is with Department of Computing, The Hong Kong Polytechnic University, Hong Kong; Email: csyqzheng@comp.polyu.edu.hk. Mo Li is with School of Computer Science and Engineering, Nanyang Technological University, Singapore; Email: limo@ntu.edu.sg. Qiuyuan Huang is with Microsoft Research, Redmond, USA; Email: qihua@microsoft.com. The authors thank Yifan Zhang for developing the smart parking application.

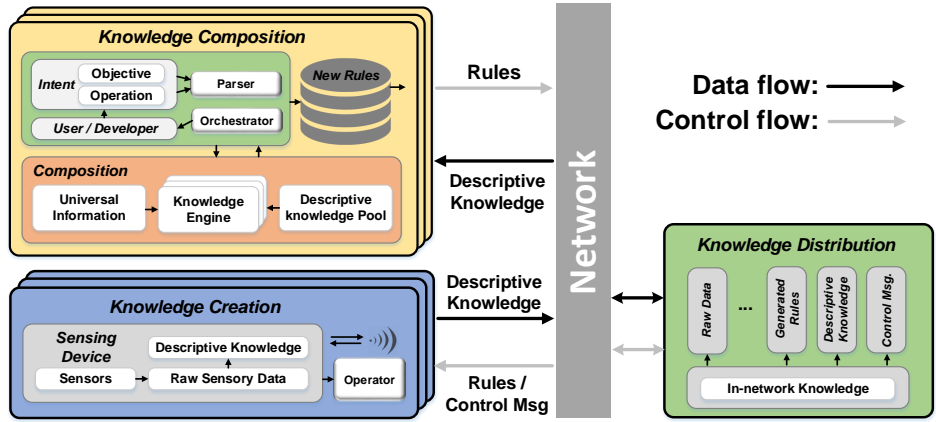

Fig. 1. Illustration of networking traffic control and system management from the KCN perspective.

to evolve automatically. These features perfectly match the complex, dynamic and time-varying nature of traffic in today's networking systems. Thus, KCN can be an emerging angle to facilitate the traffic control and networking system designs.

In fact, many initial research efforts have already been made to the three aspects above, but there still lacks a systematic review and integration of them from the KCN perspective. In particular, a relevant Knowledge-Defined Networking (KDN) concept is proposed [10], which integrates Artificial Intelligence (AI) with SDN and network analytics to further benefit and automatize network control and operation. Because the intelligence derived from KDN is mainly introduced for networking itself, which fulfills a similar functionality as KCN's knowledge distribution component. However, KCN also proposes to leverage the in-network computation and communications, *e.g.,* a complete framework from front-end sensing to back-end networking. Therefore, this paper aims to holistically review these works, present design rationale and benefits, and point out research opportunities.

## II. KCN FRAMEWORK

### A. Nutshell of KCN

Fig. 1 illustrates the architecture of the KCN framework.

1) *Knowledge creation*. IoT devices sense ambient environments and generate raw sensory data for the system. In KCN, instead of a direct injection of the raw data into the system, they are first preprocessed for extracting the **descriptive knowledge** with the immediately usable intelligence and much reduced data volume for transmission. This step is referred as the *knowledge creation* in the KCN framework.

Considering a parking space sharing system (we have its development in §III), surveillance cameras are the sensing devices. To detect whether the sparking lots are occupied and analyze the occupancy history, certain meaningful descriptors [5] will be utilized, in stead of raw video frames.

2) *Knowledge composition*. User's intent to interact with the system normally includes two perspectives: user's objective and desired operations. In KCN, based on the current objective (which may vary), how to leverage the descriptive knowledge obtained from the sensory data to automatically generate a series of in-network rules to achieve the desired objective and trigger the associated operation is referred as *knowledge composition*.

For instance, if a user wants to find a parking space near a company , the parking space sharing system needs to firstly understand the user's intent and decompose the necessary factors including current location, company's location, the estimated arrival time based on the current traffic. Hence, the relevant parking information like available time and the exact parking location is retrieved from the server. In this case, in addition to the descriptive knowledge, the composition also needs some *universal information* (human-generated rules) as a basis, *e.g.,* empty parking space can be reserved. KCN then applies learning techniques to automatically derive new rules to achieve the current objective.

Since application's objective and operation could be complicated, as shown in Fig. 1, they will be parsed first and converted to multiple decomposed elements, which are then processed by different knowledge engines. The partial results from each knowledge engine will be orchestrated as entirety to produce new rules, which will be further disseminated to the proper destinations in the system.

3) *Knowledge distribution*. Different types of information flows coexist and need to be delivered cross the network. Some flows, like the automatically generated rules, are small in size, but they normally require a higher reliability, shorter latency, higher priority in the transmission and may need to be disseminated to multiple destinations. On the contrary, the information flows, like descriptive knowledge, have relatively larger data volume, but they may tolerate certain transmission delay. In addition, different applications can further reuse the descriptive knowledge from the same sensing devices. Due to the limited networking capabilities and computing resources, how to leverage learning techniques to derive appropriate transmission rules for such a diversified traffic is named as the *knowledge distribution*.

Back to the parking example, the parking information will be returned to the user and the new generated rules will be disseminated to corresponding servers to facilitate the next searching process of this user if the system detects that this user frequently wants this parking space. As a result, when other users want a parking space in the same area at the same time as this user, other parking spaces except this one will be proposed priorly because of this rule.

So far, we briefly introduce KCN framework. Next, we will review existing works for each KCN component, and present the potential research opportunities.

*B. Knowledge creation*

Many IoT sensors could provide valuable sensing data for KCN, such as variable sensors from wireless sensor networks (WSNs), motion sensors from mobile devices, GPS on vehicles, surveillance cameras, etc. In general, there are two popular ways to create descriptive knowledge from raw sensing data: model based and deep learning based methods.

*1) Model based methods:* In this category, various model-based learning techniques, including mathematical models to describe the raw data [14] and data's inner correlation [9], are used for the knowledge creation.

Many IoT sensors are deployed to sense the ambient environments, which could provide crucial system inputs for smart urban or agriculture designs. As the varyings of many environmental factors are not arbitrary, *e.g.,* following certain patterns, one promising solution is to mathematically capture this pattern and then use the learned model to predict the future data values, instead of transmitting raw data all the time. Hence, the created descriptive knowledge is the derived mathematical model in these applications. For instance, Wang *et. al.* [11] observe that the environmental data, like temperature, may exhibit clear spatio-temporal patterns. Wu *et. al.* [14] similarly introduce a compressive sensing based model to estimate soil moisture measures. For this type of knowledge creation, the widely adopted learning techniques include linear trend analysis, compressive sensing, least-squares, etc.

The mathematical model mainly focuses on the knowledge creation for the sensors at certain given location as time goes on, *e.g.,* the temporal feature. However, sensors cannot cover every space in our daily life, *e.g.,* the sensing vacancy exists. The most representative example is the urban traffic prediction [9]. To solve such a sensing vacancy issue, one effective solution is to learn data's inner correlation from the historical data, so that the unknown sensor values can be derived by the readings from the space with the sensor coverage. In this case, the created knowledge is the inner correlation explored from the raw sensing data [9]. However, the missing values may exist in the historical data as well, which could degrade the accuracy to derive the correlation. Therefore, compressive sensing, regularized matrix factorization, low-rank approximation and interpolation can be applied to recover these missing values [15]. In summary, for this type of knowledge creation, we may need to first use numerical analysis techniques to recover missing values for the historical data and then apply learning techniques to explore data's inner correlation.

*2) Deep learning based methods:* The extracted knowledge from this category is usually the features or descriptors obtained from deep learning models [12]. One representative example is the activity recognition using the sensors from the mobile and wearable devices, such as accelerometers, gyroscopes, etc. and people need to manually define a set of features which highly impacts the system performance. Different from traditional approaches, deep learning can automatically extract a series of representative features from the input data, which is capable enough of accurately recognizing even very complicated human activities. The most widely used deep learning model is called Convolutional Neural Network (CNN). CNN contains convolution operations, which perceive valuable features from training data and pooling operations, which reduce the dimension of convolutional results and speed up entire computation for training.

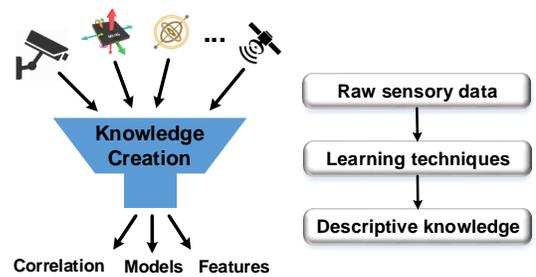

Fig. 2. Illustration of the knowledge creation in KCN.

Many research works have applied deep learning to different mobile or wearable sensory data to extract features, so that a variety of human activities can be recognized with a high precision [12]. One recent work Lasagna [8] further observes that the features obtained from different convolutional layers contain hierarchical semantics, like walking, running and jumping. Moreover, some common features can be further extracted to classify them as one category, *e.g.,* exercise. Similarly, even higher-level feature abstraction is possible to form new category, like activity. Therefore, a hierarchical understanding and searching becomes viable for mobile sensing data.

Another active area to leverage deep learning for the knowledge creation is the video processing for Compact Descriptors for Video Analysis (CDVA) [5]. As consecutive frames share many redundancies, in CDVA, certain descriptor, like scalable compressed Fisher Vector (SCFV) or color histogram, is first derived from each frame. The descriptors from two frames can be compared first. As a result, one of these two frames is directly dropped if their difference is small which indicates a high redundancy. So the traditional features can be extracted from these several selective frames merely. However, the recent standard also proposes to adopt deep learning to extract a parallel set of features, *e.g.,* Nested Invariance Pooling (NIP), which will be further integrated with the traditional feature.

*3) Edge computing facilitated knowledge creation:* However, extracting the descriptive knowledge from the raw sensory data usually requires non-trivial energy and computation resources for the sensing devices. For many embedded and miniature IoT sensors, even the generated data volume is lightweight, the devices normally do not have sufficient CPU capabilities and energies to derive the data model. In addition, the image/video data may demand more powerful CPUs to extract features or descriptors by deep learning methods.

To this end, the emerging edge computing technique serves as an promising solution to facilitate the knowledge creation. In particular, the edge computing can leverage the nearby edge servers and other available computing resources to locally and efficiently conduct the knowledge creation, *e.g.,* executing a variety of learning methods on top of the sensory data transmitted from sensing devices to unveil the hidden knowledge. As the sensory data only needs to be offloaded one-hop away to the computing edge, without traveling to the remote data center by suffering unpredictable Internet transfer delay, the cost and latency become minimum. After the descriptive knowledge is extracted, it can be rapidly delivered back from the computing edge to the sensing devices.

The edge computing essentially takes advantages of available resources nearby to provide better customized performance, which can thus significantly reduce network traffic and achieve real-time communications and lower latency. During knowledge creation, edge computing, combining with the machine learning algorithms, can also gradually learn and evolve over time, which can optimize the management of the sensory data with little human intervention.

*4) Summary:* As Fig. 2 shows, knowledge creation in KCN takes various raw sensory data as input and utilizes different learning techniques to derive descriptive knowledge, in forms of mathematical models, data's inner correlation, features, etc.

## C. Knowledge composition

Knowledge composition in KCN provides a flexible way to compose the knowledge learned (*e.g.,* descriptive knowledge) and stored in the in-network knowledge bases (*e.g.,* universal information), which can be managed by the block-chain so that every user can maintain such databases with the ensured equity and decentralization. A user's intent usually indicates what the user wants, without telling what knowledge/information the user needs or how to approach it. Continuing with the smart parking project, a user's intent could be "one parking spot near company A that I will arrive at". After receiving this intent, the parser (as in Fig. 1) could extract the following information: 1) the user's current location; 2) the user's destination; 3) the approximate arrival time according to the traffic condition on the path from the current location to the destination; 4) the user's vehicle type. After orchestrating all the knowledge together, the available parking spot information can be returned to the user.

To achieve such an intelligence, a user's intent can be expressed by any high-level declarative language such as SQL. An intent language as the North Bound (NB) Application Programming Interface (API) of Software Defined Networking (SDN) has been studied in recent years [2]. It can be served as the base of intent language for KDN. Due to the diversity nature of knowledge in KDN, the generality requirement of its intent language will be higher than that in SDN.

To understand and decompose a user's intent into the required set of knowledge calls for an intelligent parser. The parser itself requires knowledge to learn a user's intent and what the best knowledge set for such an intent is. The more intelligent the parser is, the more general the intent can be.

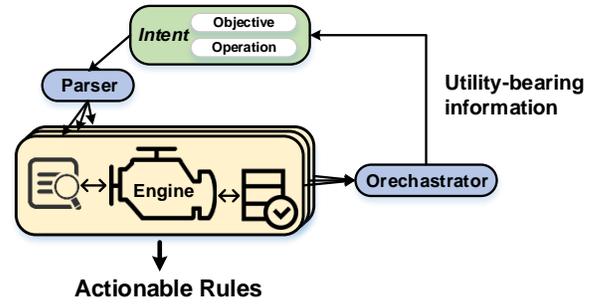

Fig. 3. Illustration of the knowledge composition in KCN.

After parsing, the parser will invoke different knowledge engines to retrieve the knowledge or take the corresponding actions in the network and/or at the edge devices.

After receiving the knowledge returned from different knowledge engines, the orchestrator will compose/fuse various knowledge together and return either composed knowledge to a user or take appropriate actions in networks. The composition/fusion is guided by various rules which are generated through human-generated knowledge or machine-generated knowledge. For instance, in case a parking spot is found, the user's current location and the traffic on its path to the destination will be fused at first to help decide the approximate arrival time. The user's destination will also be utilized to help decide which server to contact for retrieving parking spot information (assuming that the parking spot information of different areas is stored in different servers, each of which corresponds to one area). These fused pieces of information will then help decide which parking spots shall be provided to the user.

## D. Knowledge distribution

A knowledge distribution module bridges actuation/sensing devices, knowledge creation and composition modules. The key objective is to disseminate knowledge and control information to relevant devices according to users' intent and network policies. There are three scenarios for knowledge distribution: distributing knowledge to 1) devices for human consumption, 2) actuation devices for consumption of automated systems, *e.g.,* a surveillance system distributes road hazard alerts to autonomous vehicles, and 3) sensing devices for better control. To guarantee different QoS for different knowledge data packets, We promote a paradigm of SDN equipped with machine learning for intelligent control, as described next.

*1) SDN:* Different applications have different requirements on networking performance. To support these diverse requirements, the network infrastructure should allow logical abstraction and flexible reconfiguration.

Based on logical abstraction, network slicing is able to support various network functionalities and diverse performance requirements. The performance requirements include throughput, latency, coverage, priority, security, *etc*. Network slicing is a major technique in SDN and network functions visualization (NFV) [4], which enables a network infrastructure to be logically divided into multiple network service slices. Each slice provides independent logical network functions and various network resources to satisfy particular application requirements. Thus, network slicing enables concurrent deployment of multiple logically independent partitioned networks which share common physical network infrastructure. The functional separation by abstractions simplifies resource provisioning for different network services and facilitates integration of knowledge collection and knowledge dissemination. Fig. 4 illustrates the knowledge distribution in KCN. Various tasks (*e.g.,* crowdsourcing, routing) require diverse QoS performance of the network. Network slicing and virtualization optimize network resource allocation to support different network functionalities and performance requirements so that the knowledge (*e.g.,* generated rules, image/video descriptors) can be properly distributed over the network.

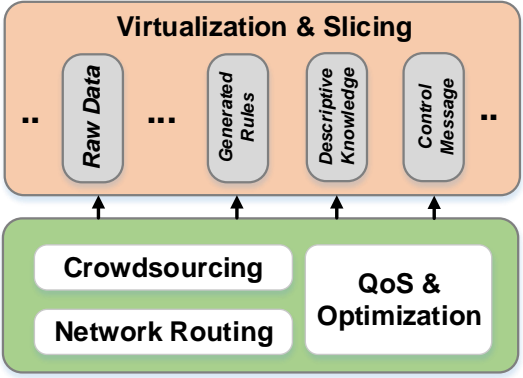

Fig. 4. Illustration of the knowledge distribution in KCN.

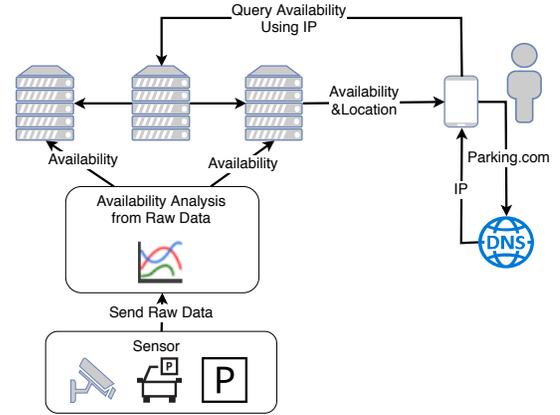

Fig. 5. Illustration of the traditional way of the application for the parking space sharing.

*2) Machine learning for intelligent control:* Machine learning algorithms implement user specified rules and optimize network resource allocation by learning to meet diverse network performance requirements of knowledge distribution. The network itself should provide interfaces and functionalities for the machine learning algorithms to flexibly work on it.

For crowdsourcing applications, the knowledge distribution module needs to value crowdsourcing tasks [6] before collecting raw data or knowledge and disseminating knowledge to actuation/sensing/display devices. An incentive mechanism should provide incentives and encourage participants to share and contribute their raw data and extracted knowledge to the network. Active crowdsourcing workers can advertise the availability and potential value of raw data labeled with its price. Meanwhile, the network should try to fetch valuable data and knowledge from many free sources (*e.g.,*, public online post) at little cost. The network then synthesizes all available knowledge and generate new knowledge. The newly discovered knowledge will be disseminated over the network and new data collection and knowledge creation tasks will be assigned to sensing devices and crowdsourcing workers.

An important problem is how to route to sensing devices to collect data and disseminate knowledge and control information over the network. The naming primitives and routing protocols of CCN can support flexible specification of knowledge of users' interest. With this capability, in an example of security monitoring and path planning, crowdsourcing workers can advertise the new data collected in dangerous areas by specifying the geographic coordinates and time stamps; the network will distribute the data with user-specified QoS; upon receiving this information, a path-planning software can find a safest route for a woman to go from her office to home (particularly useful for a woman who works at night). The network can also recruit crowdsourcing workers by publishing sensing tasks and specifying the incentives. The network can disseminate newly discovered knowledge by specifying potential values so that intended receivers can receive the knowledge with desired QoS, and exploit the knowledge for their own purposes.

In summary, machine learning algorithms take over the control of underlying network infrastructure and orchestrate network slices. Within each slice, they can dynamically reconfigure it within the performance constraints. During knowledge distribution, machine learning algorithms will gradually learn and evolve over time with little human intervention.

## III. KCN Benefits and Research opportunities

### A. Key KCN benefits

1) Although users could perform sophisticated interactions with the digital space, KCN does not require complex computations on the end devices. User's intent in KCN will be analyzed automatically. Therefore, KCN lowers the bar to encourage more end users to participate into KCN networks.

2) The data model and abstraction extracted by learning techniques can significantly reduce the data volume to be transmitted inside the network, which could maximize the networking resources and prompt the amount of concurrent users in the system simultaneously.

3) The learning techniques used in KCN could develop new networking protocols or slicing for automatically adapting to the network resource dynamics and traffic diversities. In addition, the system could further self evolve to pro-actively adjusting networking settings. Finally, the derived control logics could also guide the data content sensing to achieve a cross-layer and jointly optimized performance.

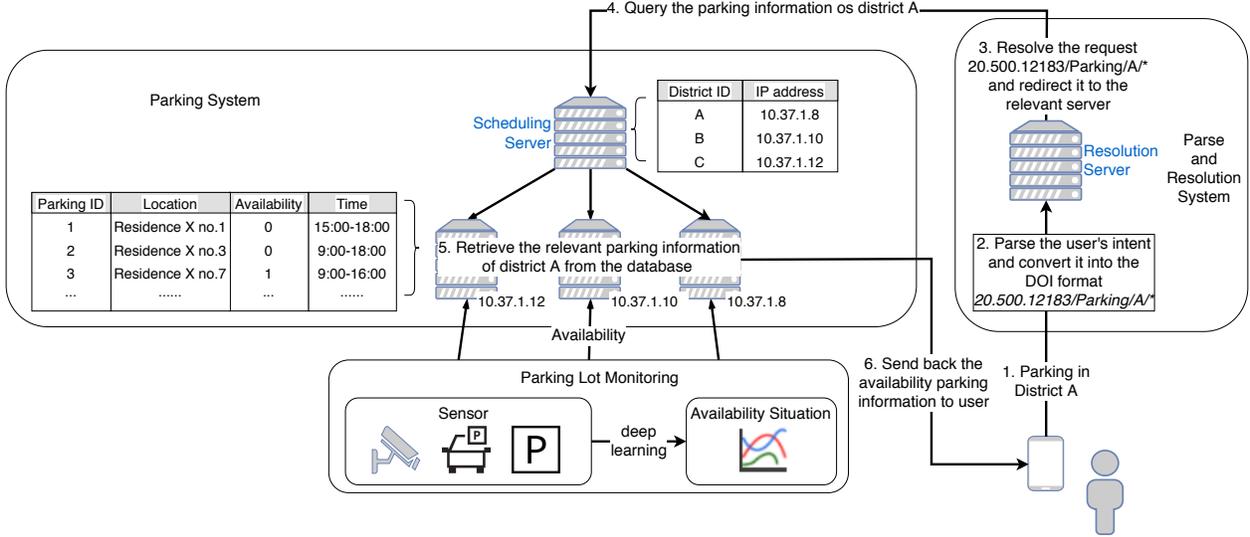

Fig. 6. Implementation of the smart parking application under the KCN framework.

*B. KCN implementation*

As a proof of concept, we implement the smart parking application under the KCN framework. Traditionally, when users want to get services from networks, the terminal first sends requests to ask for DNS resolution to get the server's IP address. Then the terminal uses this IP address to query the database at the server and acquire the wanted knowledge. To some extent, it is a host-oriented architecture as in Fig. 5.

However, KCN is a content-based architecture. We use the digital object index (*DOI*) technology, its relevant software *Handle.net* and *Handle.net* client libraries [1] to implement a KCN version in Fig. 6, where the *DOI* prefix for the parking service is *20.500.12183*. As shown in Fig. 6, in Step 1, the user places a request of *Parking in District A*. In Step 2, the user terminal parses the request and converts it into the *DOI* format. In fact, under DOI, every parking spot can be regarded as a handle. For instance, *20.500.12183/Parking/A/ResidenceX/03* refers to parking spot #03 of Residence *X* in District *A*. In Step 3, the DOI resolution server receives this request and redirects the request to the parking scheduling server whose IP address is 35.203.154.107. In Step 4, upon receiving the request, the Parking Scheduling server extracts the district information and sends this query to the servers in District A for detailed parking information. In Step 5, the servers in District A retrieve the available parking spot information from the database. In Step 6, the information is sent to the user's terminal. The user can choose one spot from the available parking spots. In addition, the occupation status of each parking spot is updated in time according to surveillance video data and the decision of users.

There are three main advantages of this KCN-based design compared with the host-oriented version: 1) With the DOI technology, the parking service can be fully decentralized. It can provide a generic and scalable platform to involve all competitors to provide parking services. Users do not need to switch among different competitors' APPs when they search for parking spots in different areas. 2) With the decentralized design, the parking spot provider's availability can be better protected. This information will not be stored in any central server, which can be potentially disclosed to a large number of users. Instead, with KCN, the availability information about a parking spot is only kept in a local server with much reduced visibility. 3) With the decentralized design, it is also easy to cope with the block-chain technologies for secure payment and privacy protection.

*C. Research opportunities*

*1) Big data collection and processing in KCN:* Machine learning techniques (*e.g.,* deep learning, deep reinforcement learning) have achieved remarkable breakthroughs. Such techniques however need a large data set to train models and extract knowledge which involves high computational overhead. Although the increasing number of computing devices would pump more raw data into the network, how to collect relevant quality data and efficiently train models remain elusive. We envision that the data collection in KCN can benefit from current naming primitives and routing protocols of CCN. Machine learning techniques can leverage the CCN to flexibly specify data of interest and collect the data set which can then serve as input data to other machine learning techniques. Distributed in-network processing promises to reduce network traffic and balance computation overhead.

*2) Knowledge management:* KCN could benefit the system management with complicated and varying user intents. However, the management of knowledge itself in KCN also calls for innovative frameworks. As knowledge in KCN can be represented in different forms, *e.g.,* models, properties of network dynamics, trained models, key parameters, how to effectively represent such knowledge so as to efficiently store, retrieve, transfer the knowledge is challenging. We envision that knowledge as an asset would have different forms to meet various application requirements. For example, a machine learning based access control scheme can exploit a trained classifier as knowledge and directly incorporate the classifier in the control scheme. The control scheme can fetch a set of trained parameters as knowledge and embed them into its feature extraction algorithm.

*3) Learning-based network optimization:* Machine learning based network performance optimization may help better allocate network resources and adapt to network dynamics. Traditional network optimization methods often monitor network metrics (*e.g.,* network throughput, latency, packet loss rates) and heuristically control networks according to handcrafted features extracted from the network metrics. By exploiting knowledge in KCN, machine generated control methods [13] may significantly improve the state of the arts and inspire network scientists and engineers to revisit traditional design principles and better understand network science and engineering. The success of KCN would extend from networks as communication infrastructure to other complex networks, *e.g.,* transportation network, social network, *etc*.

## IV. CONCLUSION

In this paper, we introduce and discuss the KCN networking paradigm that could leverage the recent success from machine learning to advance traffic control and networking system designs. We present the KCN design rationale, core components, benefits and also the research opportunities.